\documentclass{moriond}
\usepackage{xspace}

\def\be{\begin{equation}}
\def\ee{\end{equation}}
\def\bea{\begin{eqnarray}}
\def\eea{\end{eqnarray}}

\def\pt{\ensuremath{p_\mathrm{T}}\xspace}
\def\jpsi{\ensuremath{\rm J/\psi}\xspace}

\newcommand{\Photo}{\includegraphics[height=35mm]{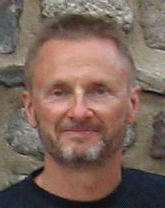}}

\begin{document}
\vspace*{4cm}
\title{Heavy-ion collision basics}

\author{ D. Mi\'skowiec }

\address{GSI Helmholtzzentrum f\"ur Schwerionenforschung GmbH, Darmstadt, Germany}

\maketitle\abstracts{ Basic concepts and terminology of relativistic
  heavy-ion collision physics are introduced and illustrated by
  experiment results. Most plots are taken from a recent ALICE
  overview paper~\cite{journey}.}

\section{Stages of a heavy-ion collision}
\begin{figure}[b]
\hspace*{9mm}\includegraphics[width=0.91\textwidth]{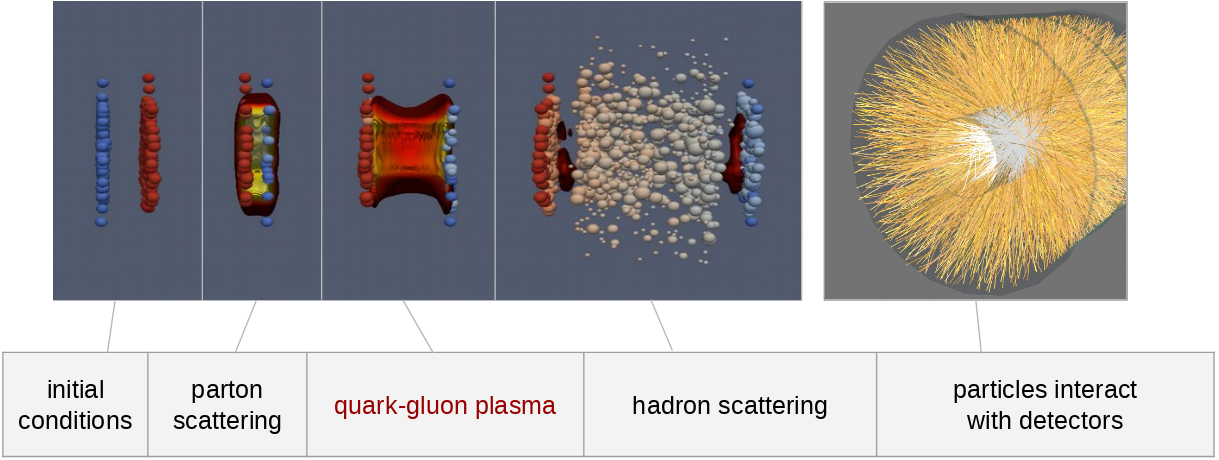}
\caption{Main stages of a Pb--Pb collision at the LHC. Sketch taken from 
Ref.~\protect\cite{madai}. See text for explanation.}
\label{fig:stages}
\end{figure}

The main stages of a Pb--Pb collision at the LHC are shown in
Fig.~\ref{fig:stages}.  
The initial state depends on the collision centrality, the
distribution of nucleons within the colliding nuclei, and the nuclear
parton distribution functions. The Lorentz-contracted colliding nuclei
pass through each other and the space between their receding disks is
filled with energy in form of gluons and quarks (quark-gluon
plasma). It is convenient to consider collisions between pairs of
nucleons~\cite{czyz}.  The temporal sequence of these collisions does
not depend on their longitudinal positions within nuclei. Instead, all
nucleon--nucleon (NN) collisions are initiated at the same time and
their duration depends on their hardness.  Hard collisions finish
first and get all the energy they want; this is why they scale with
the number of NN collisions $N_{\rm coll}$.  Soft collisions take
longer and compete among themselves for energy, so soft particles
scale with the total available energy or the number of participating
nucleons, $N_{\rm part}$.  Hard collisions do not compete against soft
collisions because they are faster, and do not compete among
themselves because they are rare.  Heavy quarks and high-\pt~quarks
and gluons originate from the initial hard processes. During
subsequent reaction stages they interact with the bulk matter probing
its properties.  The fireball keeps expanding until the energy density
drops below 1~GeV/fm$^3$, at which point the quarks and gluons have to
turn into color-neutral hadrons.  These still interact with each other
for some time, then propagate freely to the detectors.

\section{Initial conditions}
\begin{figure}[h]
\begin{minipage}{0.55\textwidth}
\hspace*{0mm}\includegraphics[height=5.7cm]{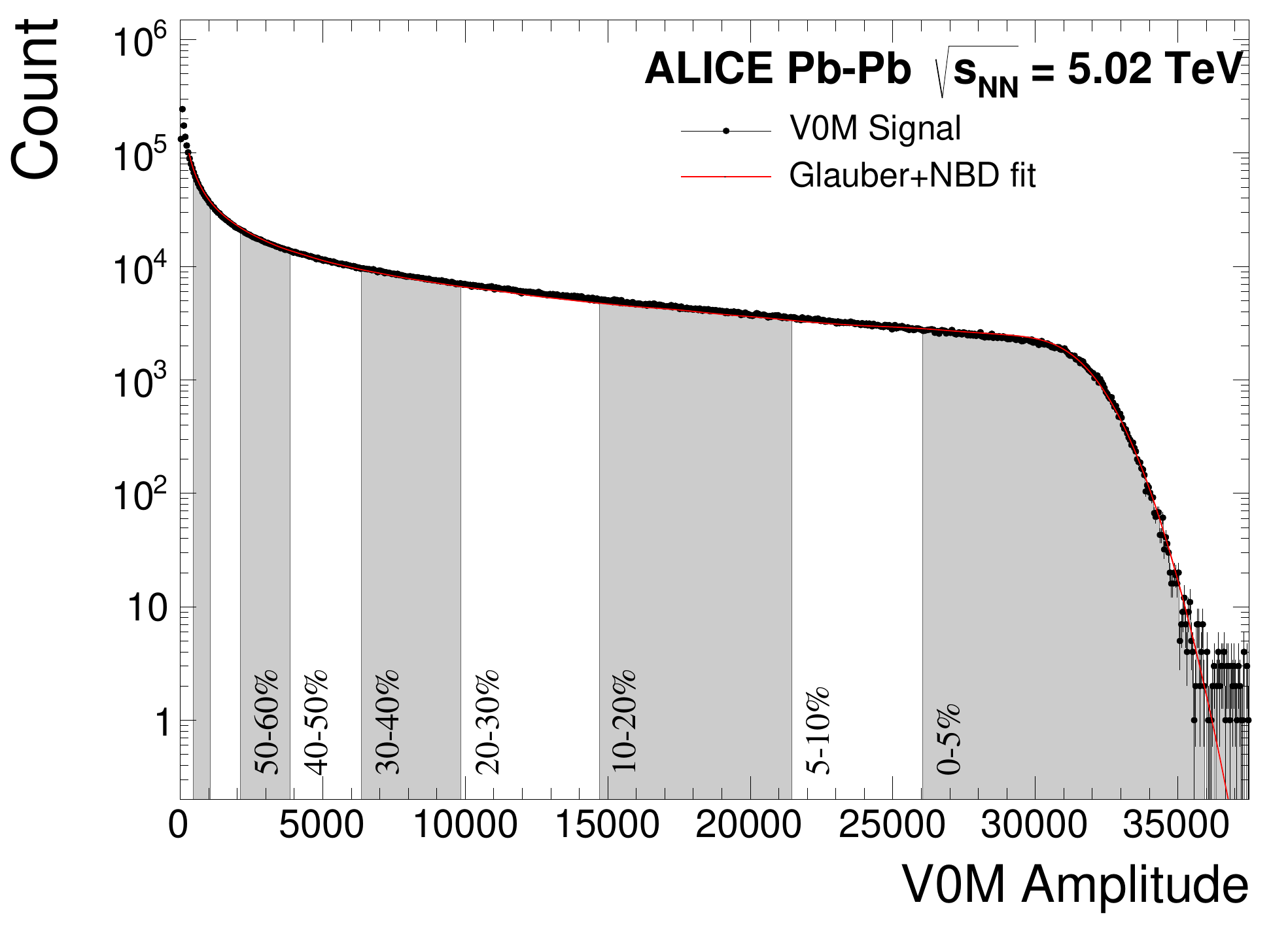}
\end{minipage}
\begin{minipage}{0.45\textwidth}
\includegraphics[height=4.8cm]{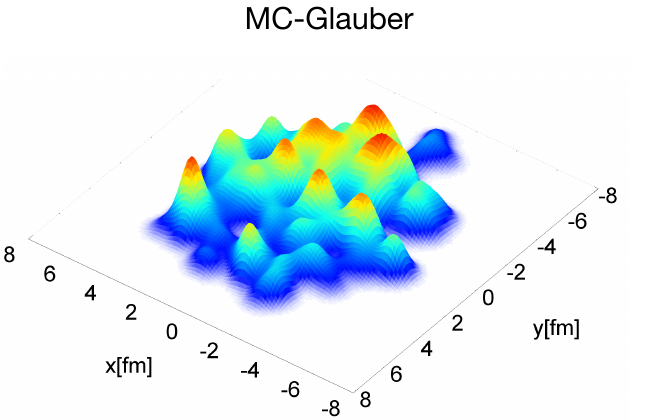}
\end{minipage}
\caption{Initial conditions of a relativistic nucleus--nucleus collision:  
nucleus--nucleus centrality (left) and the transverse distribution of 
energy as resulting from nucleon positions in the nuclei (right). Plots from Ref.~\protect\cite{journey}.}
\label{fig:initial12}
\end{figure}

\begin{figure}[b]
\begin{minipage}{0.59\textwidth}
\includegraphics[height=5.25cm]{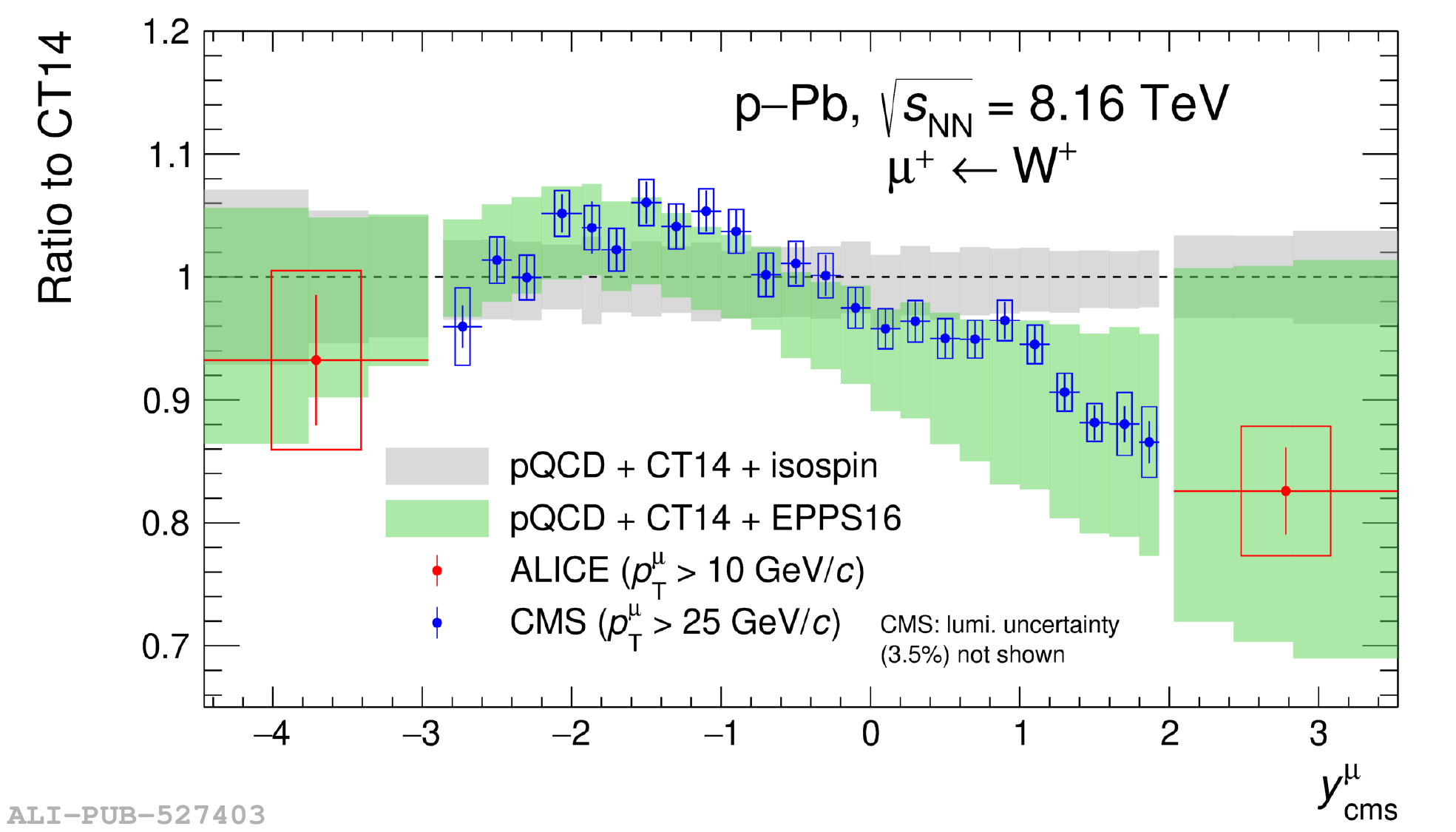}
\end{minipage}
\begin{minipage}{0.41\textwidth}
  \caption{CMS and ALICE data on W-boson production in p--Pb collisions
    at the LHC~\protect\cite{journey}. This measurement is sensitive
    to the parton distribution function in Pb nucleons.}
\label{fig:initial3}
\end{minipage}
\end{figure}

The initial state depends on the collision centrality, the transverse
distribution of nucleons within the colliding nuclei, and the nuclear
parton distribution functions.  Knowing the nuclear collision cross
section d$\sigma$/d$b = 2 \pi b$ (for $b$ smaller than the sum of the
radii of the two nuclei) and assuming that the multiplicity of the
produced particles decreases with increasing $b$, the centrality can
be estimated from the particle multiplicity or any observable
proportional to it.  Its differential cross section has to be
integrated from the right and normalized to the known total collision
cross section. As an example, events with centrality 0--5\% in ALICE
are selected by requiring a particular detector (V0M) signal to be
above $26\cdot 10^3$ (Fig.~\ref{fig:initial12} left).  This centrality
corresponds~\cite{overlap} to 0$\,<b<\,$3.5~fm.  For a fixed $b$, the
fireball shape will still depend on the transverse distribution of
energy as resulting from the nucleon positions prior to the collision
(Fig.~\ref{fig:initial12} right).
The third ingredient of the initial conditions are the parton
distribution functions in nucleons, which differ from those of free
protons (shadowing). They can be studied, in the relevant
Bjorken-$x$ range, e.g. by measuring W production in proton--nucleus collisions
(Fig.~\ref{fig:initial3}).

\section{Parton scattering}
\begin{figure}[h]
\center{\includegraphics[width=0.5\textwidth]{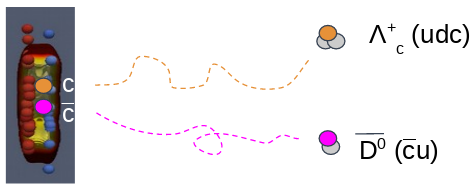}}
\caption{Charm quarks are produced in the initial parton scattering, 
interact with the expanding QGP, and hadronize into charmed mesons 
and baryons. }
\label{fig:heavy-quarks}
\end{figure}

Heavy quarks are produced in q$\bar{\rm q}$ pairs in the initial
parton scatterings with high momentum transfer. Their final fate --
the hadrons in which they end up and the momenta thereof -- depend on
the subsequent stages of the system evolution, in particular on their
interaction with the quark-gluon plasma and, apparently, on the
environment in which they hadronize (Fig.~\ref{fig:heavy-quarks}).

A standard way of quantifying the interaction of heavy quarks with the
QGP consists in comparing the transverse momentum spectra of hadrons
carrying these quarks in nucleus--nucleus (AA) and pp collisions. The
nuclear modification factor is defined as \be R_{\rm AA} (\pt) =
\frac{1}{\langle N_{\rm coll} \rangle} \frac{{\rm d}N_{\rm AA}/{\rm
    d}\pt}{{\rm d}N_{\rm pp}/{\rm d}\pt} \ee with
$\langle N_{\rm coll} \rangle$ being the number of nucleon--nucleon
collisions within the nucleus--nucleus collision, averaged over the
centrality bin.  Since hard processes scale with the number of
nucleon--nucleon collisions, $R_{\rm AA}$ would be unity for
(hypothetical) non-interacting heavy quarks (except for shadowing), as
it is indeed the case for hard direct photons and W bosons.

\section{Quark-gluon plasma}
\begin{figure}[h]
\center{\includegraphics[width=0.7\textwidth]{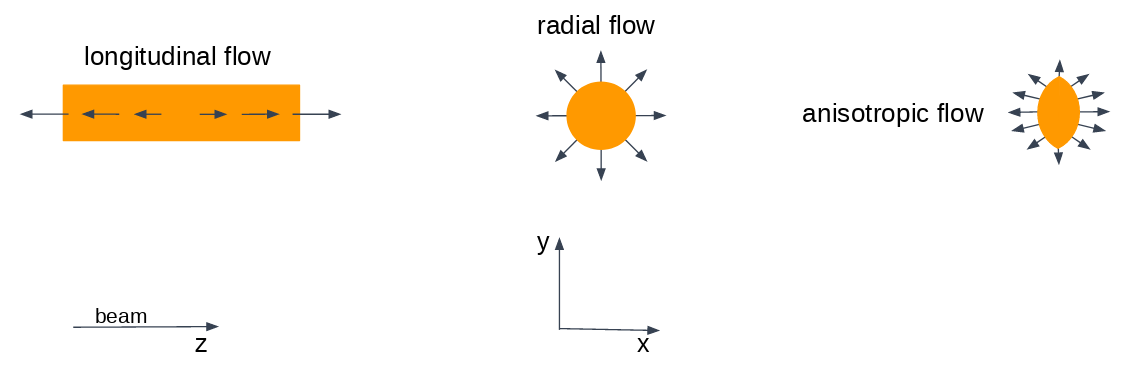}}
\caption{Longitudinal and transverse flow in relativistic heavy-ion 
collisions.}
\label{fig:flow}
\end{figure}

Before one can study the interaction of heavy quarks with the QGP, the
geometry and expansion of the QGP itself must be known. The collective
longitudinal and transverse expansion (aka flow) of the QGP are here
the main features. ``Collectivity'' can be defined as the lack of
independence between the positions and velocities of the
particles/fluid elements.  The longitudinal expansion results mainly
from the dependence of particle position (at a finite time) on the
particle velocity (at the time of the full overlap of the two
colliding nuclei). The transverse expansion, on the other hand, is
driven by the preferred direction of emission of particles away from
the opaque high-density zone. In non-central collisions, the
transverse expansion may depend on the azimuthal angle.  These modes
of flow, sketched in Fig.~\ref{fig:flow}, influence the momenta of
emitted particles.

\begin{figure}[h]
\center{\includegraphics[width=0.49\textwidth]{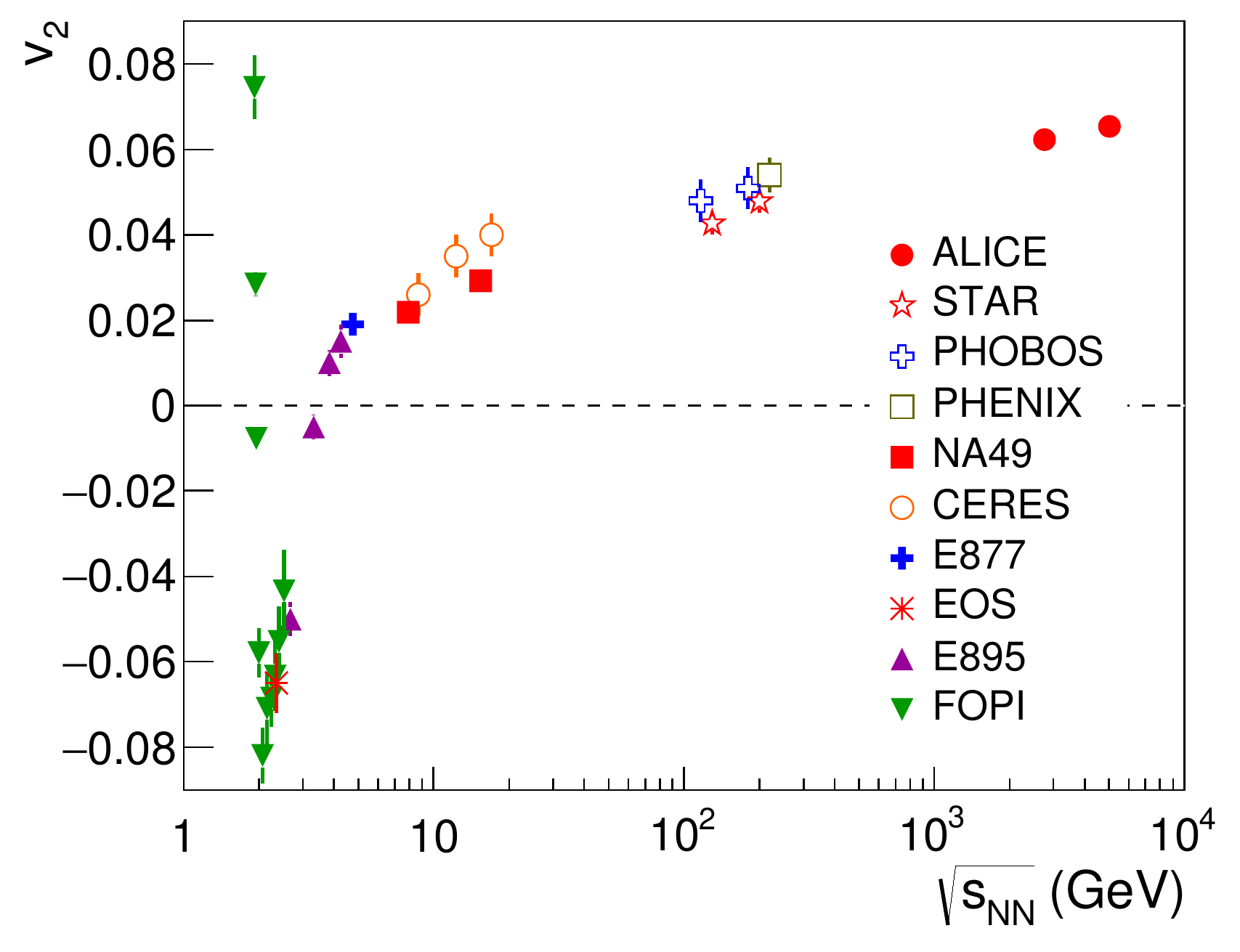}}
\vspace{-2mm}
\caption{Collision energy dependence of the elliptic flow in 
midcentral (20--30\%) lead and gold collisions~\protect\cite{journey}.}
\label{fig:v2}
\end{figure}

The elliptic flow, quantified by the second Fourier harmonic of the
azimuthal distribution of particles with respect to the reaction
plane, is a particularly interesting observable. The reaction plane is
defined by the beam axis and the impact parameter. The collision
energy dependence of the elliptic flow parameter $v_2$ in midcentral
(20--30\%) lead and gold collisions is shown in Fig.~\ref{fig:v2}.
The preferred emission of particles at the lowest and highest energy
is in plane, while at 1.5--3 GeV the in-plane emission is obstructed by
the presence of the non-overlapping parts of the colliding nuclei
(spectators).

The high values of $v_2$ observed at 130--200~GeV at RHIC were in
agreement with hydrodynamic calculations, indicating that the QGP
behaved like a non-viscous liquid rather than a gas of non-interacting
quarks and gluons as had been initially expected. Viscosity would
smear out the velocity fields and thus would reduce $v_2$, this was
not observed. A big question was whether $v_2$ would be lower at the
LHC. It turned out that the QGP there was as liquid and as perfect as
at RHIC; the $v_2$ is even slightly higher, but this is only because
of its scaling with \pt.

\begin{figure}[bh]
\begin{minipage}{0.57\textwidth}
\includegraphics[height=6cm]{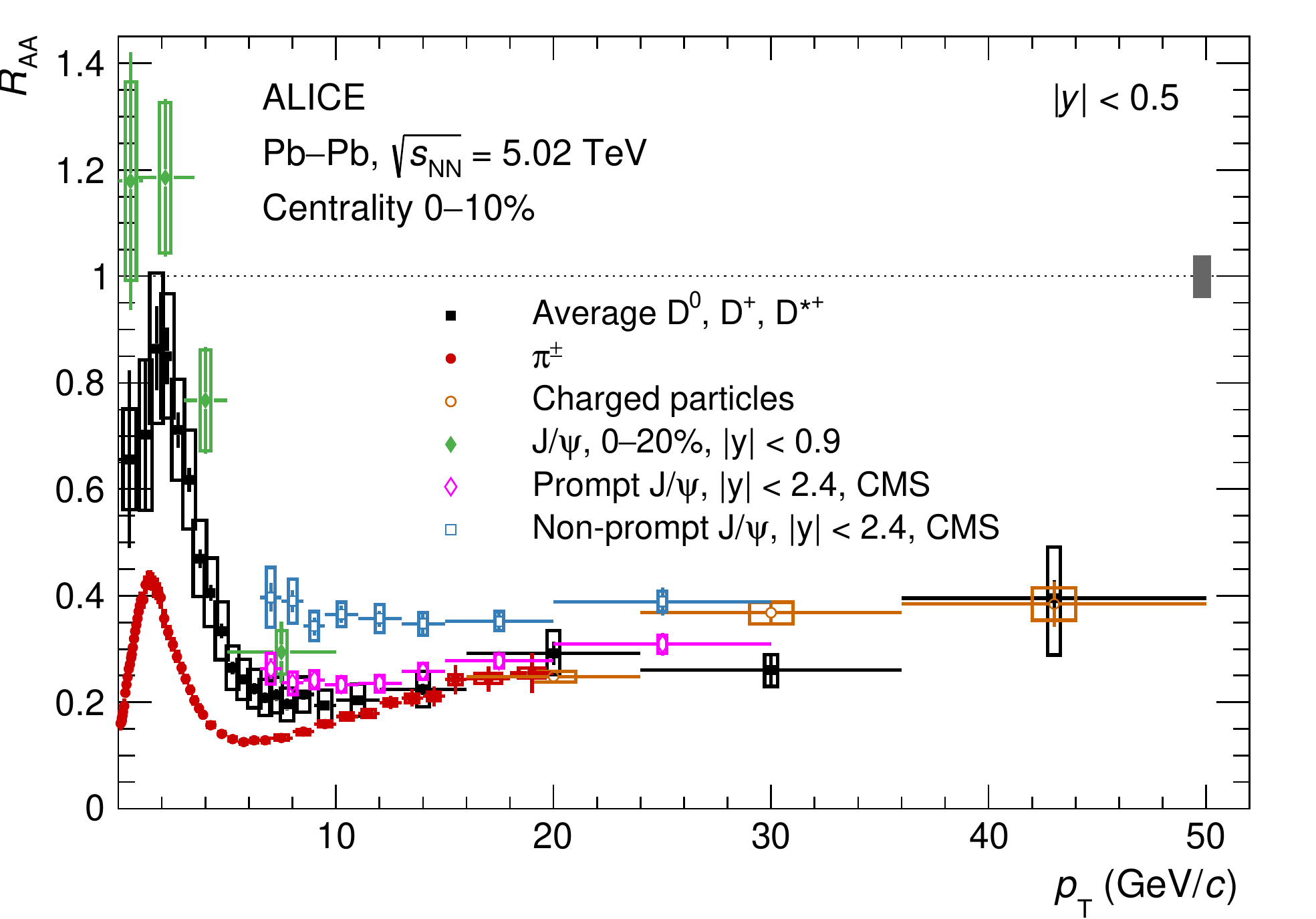}
\end{minipage}
\begin{minipage}{0.43\textwidth}
\vspace*{2mm}\includegraphics[height=6cm]{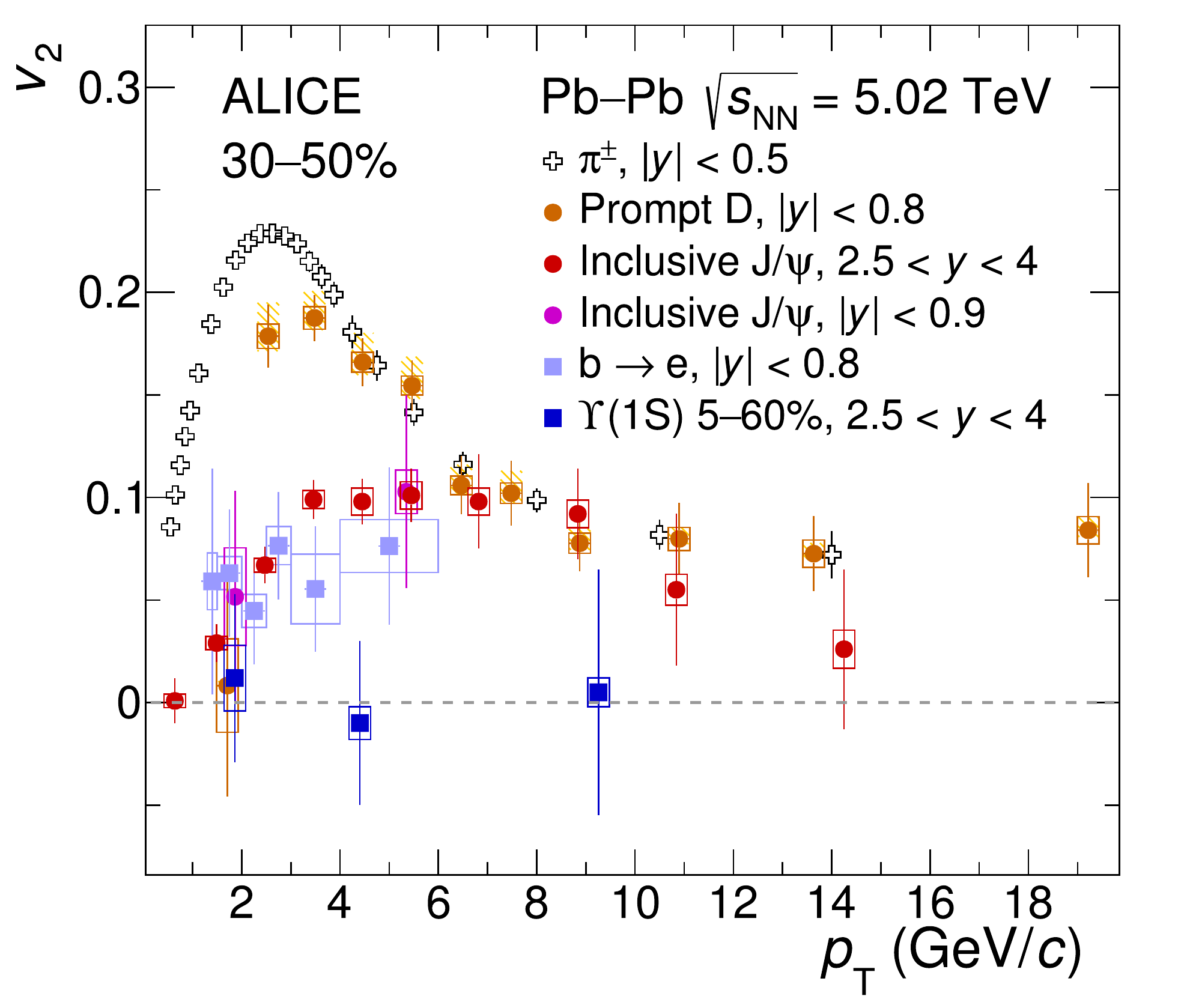}
\end{minipage}
\caption{Nuclear modification factor $R_{\rm AA}$ (left) and elliptic
  flow parameter $v_2$ (right) of various particle species. Non-prompt
  \jpsi mesons come from beauty decays and thus are a proxy for b
  quarks. The left and right plots are taken from
  Refs.~\protect\cite{charmpaper} and \protect\cite{journey},
  respectively.}
\label{fig:hf}
\end{figure}

Heavy quarks and high-momentum particles are important probes of the
QGP.  Energetic (faster than bulk) heavy quarks have to traverse the
QGP before they can leave the fireball. The energy loss they
experience, combined with the steeply falling hadron \pt spectra,
leads to a reduced value of $R_{\rm AA}$ (Fig.~\ref{fig:hf} left).
Slight differences between the particle species (mass ordering)
reflect the radiative energy loss differences related to the dead 
cone effect~\cite{deadcone}.
The interaction of heavy quarks with the QGP and the fact that the
final charm and beauty hadrons contain also light quarks lead to
these hadrons showing an azimuthal anisotropy of emission similar to
light hadrons, albeit reduced in strength (Fig.~\ref{fig:hf} right). 

\begin{figure}[h]
\begin{minipage}{0.55\textwidth}
\vspace*{-1cm}\hspace*{0mm}\includegraphics[height=4cm]{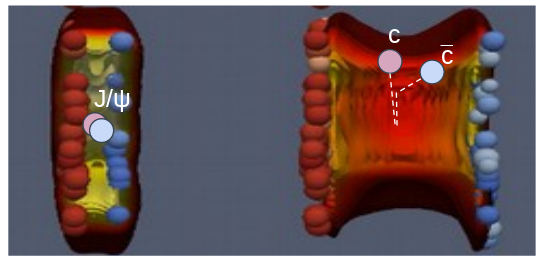}
\end{minipage}
\begin{minipage}{0.45\textwidth}
\hspace*{5mm}\includegraphics[height=6.5cm]{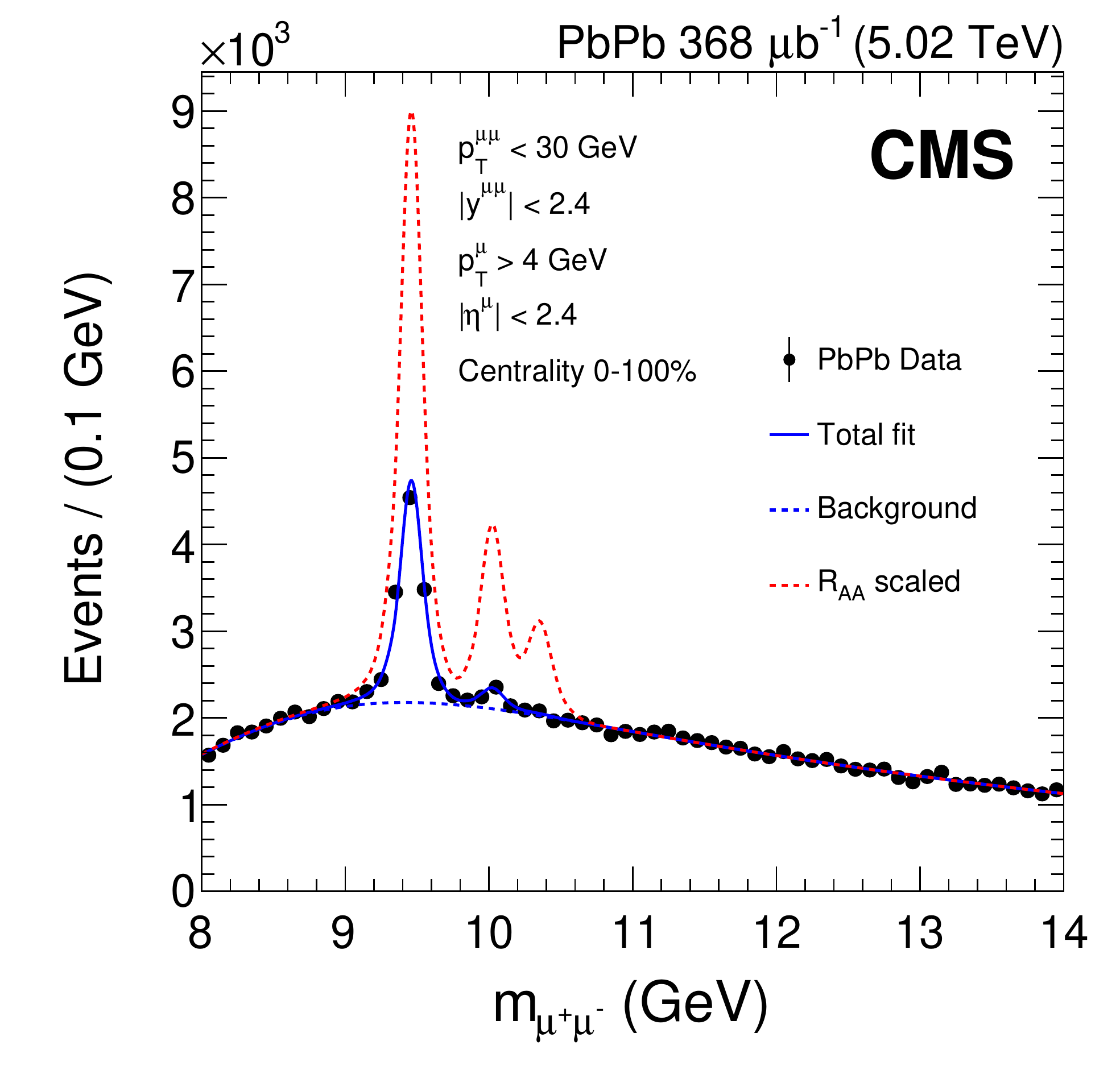}
\end{minipage}
\vspace{-5mm}
\caption{Left: \jpsi creation in the initial parton scattering and
  possible dissociation in the QGP.  Right: Suppression of
  $\Upsilon$(1S), $\Upsilon$(2S), and $\Upsilon$(3S) in Pb--Pb
  collisions compared to pp (figure taken from
  Ref.~\protect\cite{cmsups}).}
\label{fig:jpsi-melting}
\end{figure}

The suppression of charmonia and bottomonia, and especially of their
higher states, is another important signature of the presence of the
QGP.  Quarkonia, if at all formed, can dissociate into c and
$\bar{\rm c}$ (or b and $\bar{\rm b}$) by the interaction with the QGP
before they have a chance of leaving the fireball
(Fig.~\ref{fig:jpsi-melting} left). The suppression of bottomonium in
Pb--Pb collisions at the LHC is clearly seen in the CMS data shown in
the right panel of Fig.~\ref{fig:jpsi-melting}.  \vspace{-3mm}

\section{Hadronization}
The critical temperature at which the QGP turns back into hadrons is
155 MeV~\cite{bazavov,katz}.  At the LHC the standard vacuum
fragmentation process known from e$^+$e$^-$ colliders seems to be
augmented by another hadronization mechanism, quark
coalescence. Indications for this can be found in the \jpsi yield and
in the charm fragmentation fractions.
\begin{figure}[b!]
\begin{minipage}{0.6\textwidth}
\includegraphics[height=6.2cm]{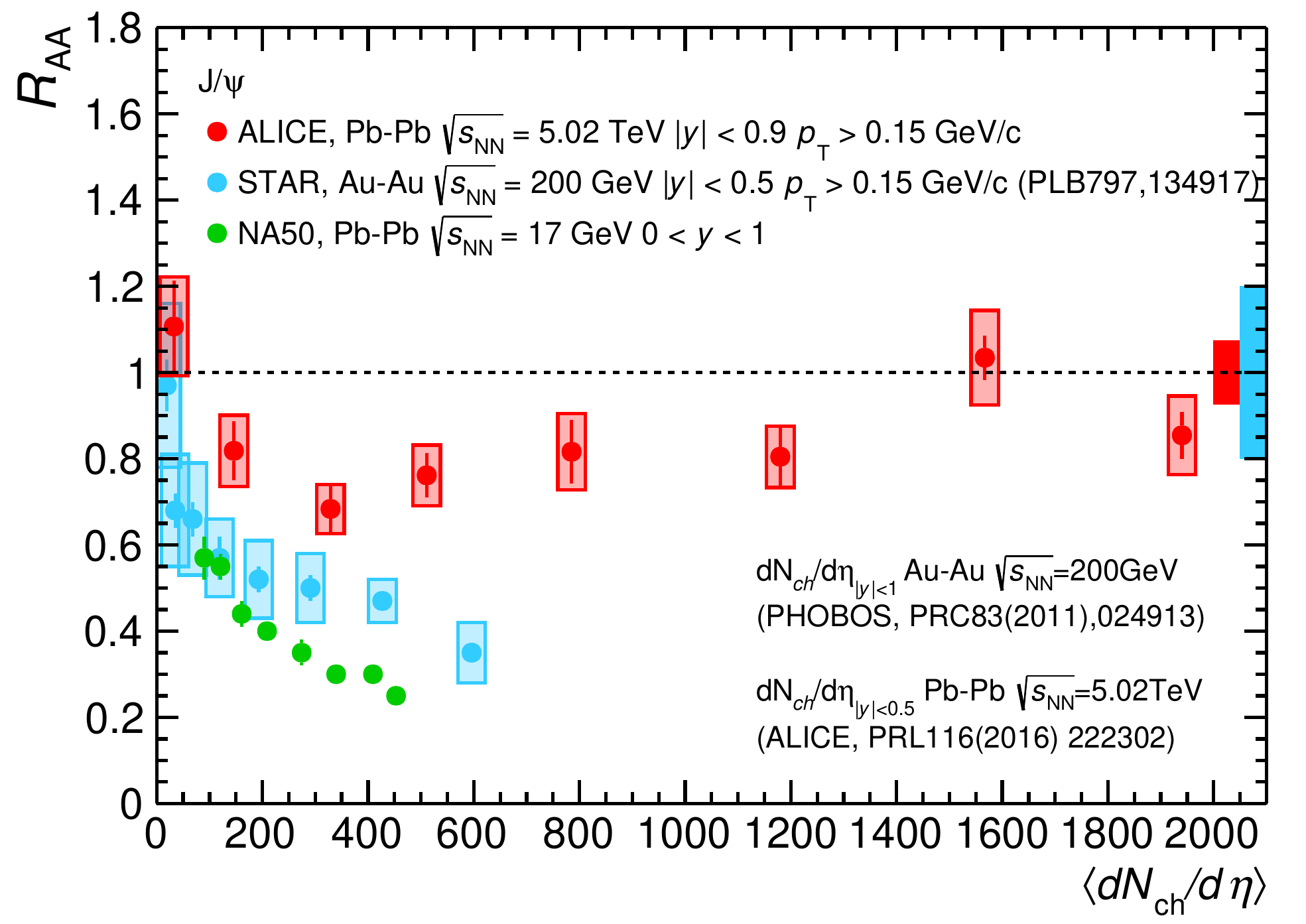}
\end{minipage}
\begin{minipage}{0.4\textwidth}
\includegraphics[height=6cm]{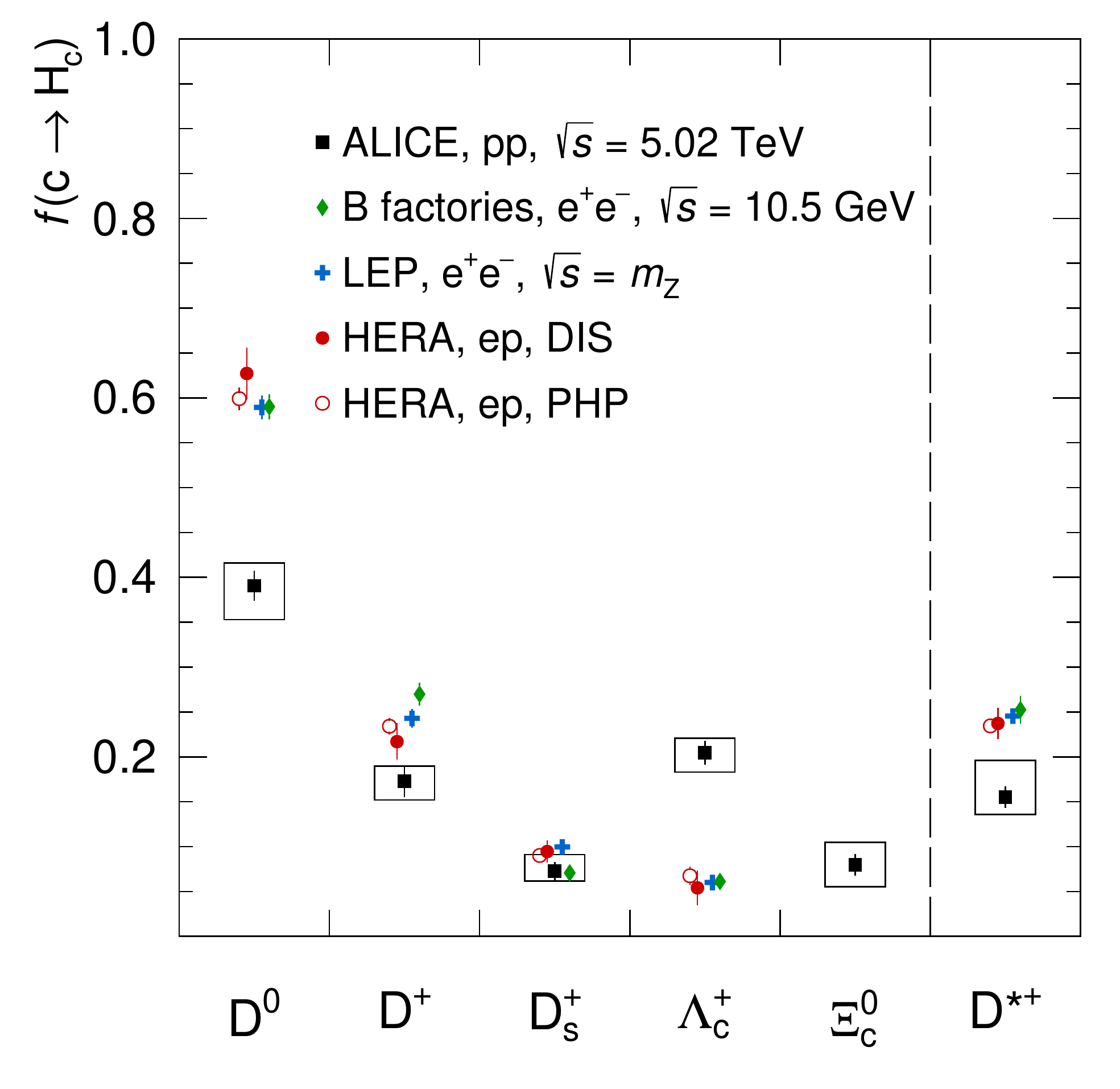}\vspace{-3mm}
\hspace*{11mm}\includegraphics[height=3.9mm]{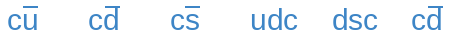}
\end{minipage}
\caption{Left: Nuclear modification factor of
  \jpsi~\protect\cite{journey}. In the most central Pb--Pb collisions,
  the suppression is compensated by an additional production channel
  of \jpsi via c--$\bar{\rm c}$ (re)combination.  Right: fragmentation
  fractions of c quarks in pp
  collisions~\protect\cite{charm-hadronization}. The baryon/meson
  ratio is enhanced in pp compared to ee and ep collisions.}
\label{fig:hadr}
\end{figure}
With about 15 c$\bar{\rm c}$ pairs produced per unit rapidity, the
predicted~\cite{pbm} additional production of \jpsi via
c--$\bar{\rm c}$ (re)combination (or \jpsi (re)generation) seems
indeed to enhance the low-\pt \jpsi yield in the most central Pb--Pb
collisions (Fig.~\ref{fig:hadr} left).  Another hint for quark
coalescence can be found in the charm fragmentation fractions. The
relative abundance of $\Lambda_{\rm c}$ in pp is significantly
enhanced -- and D$^0$ reduced -- compared to the ee and ep collision
experiments (Fig.~\ref{fig:hadr} right).

\section{Hadron scattering}
\begin{figure}[h]
\center{\includegraphics[width=0.93\textwidth,clip,trim = 0cm 55mm 0cm 0.8cm]{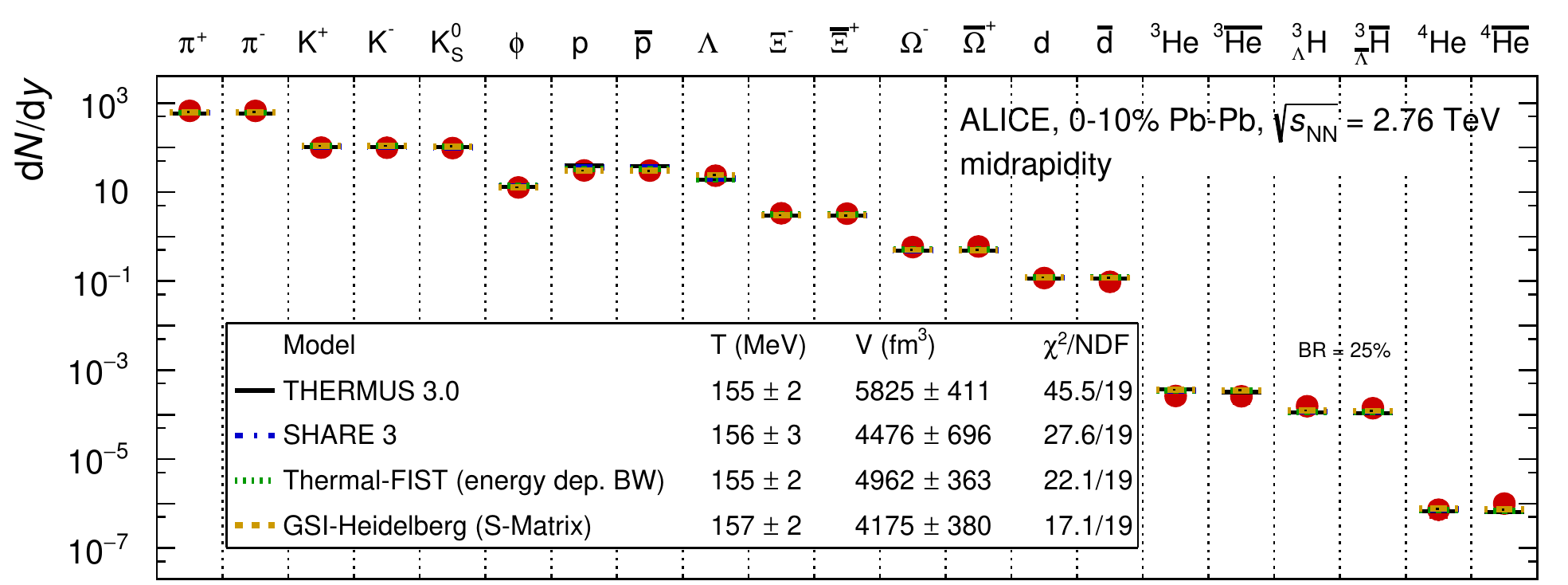}}
\caption{Hadron yields and their description by the statistical
  model~\protect\cite{journey}. The extracted chemical freezeout
  temperature is 155--157 MeV, similar to the critical temperature at
  which the QGP turns into hadrons.}
\label{fig:tchem}
\end{figure}

The freshly formed hadrons may scatter before the density drops enough
to allow for their free propagation to detectors. The yields of hadron
species get fixed when the inelastic interactions cease (chemical
freezeout).  Similarly, the end of elastic interactions (kinetic
freezeout) shapes the transverse momentum spectra of hadrons. The
chemical freezeout temperature, extracted from the hadron yields
(Fig.~\ref{fig:tchem}) is about 156~MeV, thus coinciding with the
critical temperature at which the hadronization takes place.

The kinetic freezeout is explored by simultaneously fitting the shapes
of the transverse momentum spectra of various hadrons with just two
parameters representing the temperature and the transverse flow
velocity.  This so-called ``blast-wave fit'' yields temperatures
between 100 and 160~MeV depending on the treatment of resonances which
feed into light mesons, and thus leaving -- or not -- room for the
hadronic scattering stage in the evolution of the
fireball~\cite{journey}. The presence of a hadronic stage is supported
e.g. by the observed reduction of K$^*$ ($c\tau=4.2$~fm) yield
in central Pb--Pb collisions, presumably caused by scattering
of its decay daughter pion.

\section{Conclusions}
While the main object of study in relativistic heavy-ion collisions is
the quark-gluon plasma, a detailed understanding of all stages of the
reaction is necessary for proper interpretation of the measurements.
The author thanks for the invitation to Moriond and acknowledges the
pleasant and inspiring atmosphere of the meeting.

\end{document}